\documentclass[prl,twocolumn,showpacs]{revtex4}
\usepackage{graphicx}
\usepackage{amsmath,amssymb}
\usepackage{bm}

\begin{document}

\title{Isotropic magnetometry with simultaneous excitation of orientation and
alignment CPT resonances}
\author{A. Ben-Kish, M. V. Romalis}
\affiliation{Department of Physics, Princeton University, Princeton,
New Jersey 08544, USA}

\date{\today}

\begin{abstract}
Atomic magnetometers have very high absolute precision and
sensitivity to magnetic fields but suffer from a fundamental
problem: the vectorial or tensorial interaction of light with atoms
leads to "dead zones", certain orientations of magnetic field where
the magnetometer loses its sensitivity. We demonstrate a simple
polarization modulation scheme that simultaneously creates coherent
population trapping (CPT) in orientation and alignment, thereby
eliminating dead zones. Using $^{87}$Rb in a 10 Torr buffer gas cell
we measure narrow, high-contrast CPT transparency peaks in all
orientations and also show absence of systematic effects associated
with non-linear Zeeman splitting.
\end{abstract}

 \maketitle


Atomic magnetometers measure the magnetic field by detecting the
Zeeman spin precession frequency and can achieve sensitivity
surpassing even the best superconducting quantum interference
devices (SQUID) \cite{Budker_Romalis_Nature_Phys}. They are
particularly suitable for operation in Earth's magnetic field
because measurements of the spin precession frequency can be
performed with high fractional resolution, are related to the
magnetic field only through fundamental constants, and are
relatively insensitive to the orientation of the magnetometer. As a
result, atomic magnetometers are widely used for the most demanding
applications, such as measurements of magnetic fields in space
\cite{Cassini,Juno,Swarm}, mineral exploration
\cite{Nabighian_2005}, searches for archeological artifacts
\cite{David_2004}, and unexploded ordnance \cite{Nelson}.

Since their inception, all types of atomic magnetometers have
suffered from a fundamental problem known as a "dead zone"
\cite{Bloom_APL_1962}: for certain orientation of the magnetic field
relative to the device the magnetometer signal goes to zero. Dead
zones are an inherent feature of the vector or tensor interactions
used for optical pumping and detection of spin oscillations: for
certain orientations the interaction term goes to zero. Previous
solutions to this problem included using multiple magnetometer cells
or beams \cite{Cheron_J_Phys_III_1997,Cheron2000}, mechanical
rotation of some components\cite{Guttin_J_De_Phys_1997}, and use of
unpolarized light and spatially-varying microwave fields
\cite{Aleksandrov2006}. Particularly for space magnetometers it
presents a serious challenge that requires increased complexity of
the system \cite{Juno,Swarm}.

Here we demonstrate a dead-zone-free Rb magnetometer utilizing a
single interaction region. The magnetometer uses a single laser beam
with polarization modulation at the Zeeman frequency to
simultaneously excite and detect the precession of the dipole and
quadrupole (alignment and orientation) moments of the atomic density
matrix. Because the vector nature of the two interactions is
different, the transmission of the laser beam is sensitive to the
magnetic field for all orientations.

Excitation of magnetic resonances using light intensity modulation
was first demonstrated by Bell and Bloom \cite{Bell_Bloom_1957} and
polarization modulation was explored in
\cite{Gilles_Opt_Comm_1991,Suter}. Magnetometery based on precession
of alignment was extensively explored in non-linear magneto-optical
rotation (NMOR) magnetometers \cite{Buder_RMP_2002_NMOR} while
transmission monitoring has been used in CPT magnetometers
\cite{Wynands_1998,Wynands2001,Kitching}. However, in all these
cases the magnetometer signals go to zero for certain orientations
of the magnetic field. We show that with a particular choice of
light modulation parameters and the detection method one can realize
a CPT magnetometer operating simultaneously on two different
coherences without any active adjustments, resulting in a large
response for all orientations of the magnetic field.

In addition to avoiding dead zones our arrangement also largely
eliminates ''heading errors'', another long-standing problem,
particulary for alkali-metal magnetometers. Heading errors are due
to non-linear Zeeman splitting of the magnetic sublevels given by
the Breit-Rabi formula. By
using symmetric optical pumping with equal intensities of $\sigma ^{+}$ and $%
\sigma ^{-}$ light we eliminate spin polarization along the magnetic
field. The remaining third-order heading error \cite{Seltzer_2007}
is suppressed by the square of the ratio of the Zeeman frequency to
the hyperfine frequency. The error due to nuclear magnetic moment is
also avoided by optical pumping on only one of the hyperfine ground
states.

Consider light polarized linearly in the $\hat{x}$ direction with
intensity $I$ propagating parallel to the $\hat{z}$ axis, passing
through a polarization modulator with an optic axis at 45$^{\circ}$
to $\hat{x}$ and a retardation angle $\phi =\phi _{0}\cos \omega t$.
The following Stokes parameters of the light will be modulated
\begin{eqnarray}
S_{x} &=&I\cos (\phi _{0}\cos \omega t), \\
S_{z} &=&I\sin (\phi _{0}\cos \omega t).
\end{eqnarray}
The interaction of light with atoms can be written as an effective
non-hermitian ground-state Hamiltonian \cite{Happer_RMP_1972}
\begin{equation}
\left\langle n|H|m\right\rangle =\sum_{m^{^{\prime }}}\frac{\mathbf{E}%
^{*}\cdot \mathbf{d}_{nm^{\prime }}\mathbf{d}_{m^{\prime }m}\mathbf{\cdot E}%
\,}{\hbar (\omega _{0}-\omega _{F,F^{\prime }}+i\Gamma /2)},
\end{equation}
where $\mathbf{E}$ and $\mathbf{d}$ are the electric field and
electric dipole operator, $\omega _{0}$ is the laser frequency,
$\omega _{F,F^{\prime }}$ is the transition frequency from the
ground state $F$ to the excited state $F^{\prime }$, and $\Gamma $
is the excited state decay rate. The Hamiltonian can be decomposed
into a sum of scalar, vector and tensor
components\cite{Happer_RMP_1972,Mabuchi_2006}.  Retaining only terms
proportional to the modulated Stokes parameters we obtain
\begin{equation}
H_{mod}=\frac{2}{c\epsilon}\sum_{F,F^{\prime }}\frac{\alpha
_{F,F^{\prime }}^{(1)}S_{z}F_{z}+\alpha _{F,F^{\prime
}}^{(2)}S_{x}(F_{x}^{2}-F_{y}^{2})}{\hbar(\omega _{0}-\omega
_{F,F^{\prime }}+i\Gamma /2)} , \label{eq:Hamiltonian}
\end{equation}
where  $\alpha _{F,F^{\prime }}^{(1)}$,  $\alpha _{F,F^{\prime
}}^{(2)}$ are the vector and tensor atomic polarizability constants
respectively, defined in \cite{Mabuchi_2006}. Light absorption in
the vapor and the  de-population optical pumping rate are both
proportional to the non-hermition part of the Hamiltonian $H-H^{\dag
}$
\cite{HapperMathurPR1967,MathurTangHapperPRA1970,Happer_RMP_1972}.
In
particular, for weak light intensity the atoms will develop orientation $%
\left\langle F_{z}\right\rangle \propto -S_{z}$ and alignment
$\left\langle F_{x}^{2}-F_{y}^{2}\right\rangle \propto -S_{x}$ due
to \emph{depopulation} optical pumping. The effects of
\emph{repopulation} optical pumping are  relatively small in the
presence of buffer gas due to fast excited state spin relaxation.

The modulation of the Stokes parameters can be expanded in terms of
Bessel functions:
\begin{eqnarray}
S_{x} &=&J_{0}(\phi _{0})+2\sum_{k=1}^{\infty }(-1)^{k}J_{2k}(\phi
_{0})\cos 2k\omega t \\
S_{z} &=&2\sum_{k=0}^{\infty }(-1)^{k}J_{2k+1}(\phi _{0})\cos
(2k+1)\omega t
\end{eqnarray}
One can see that for small $\phi _{0}$ we get modulation of $S_{z}$ at $%
\omega $ and modulation of $S_{x}$ at $2\omega $. The Bell-Bloom
(BB) magnetometer \cite{Bell_Bloom_1957} (Fig. 1A) relies on the
Larmor precession of the atomic orientation $\left\langle
F_{z}\right\rangle.$ If the frequency of the modulation $\omega $
matches the Larmor precession frequency $\omega _{L}$, then
$\left\langle F_{z}\right\rangle $ will be synchronously excited due
to depopulation pumping by the modulation of $S_{z}.$ The average
transmission of the light through the cell will be increased due to
the term $S_{z}F_{z}\propto -\cos ^{2}(\omega t)$. However, when the
magnetic field is parallel to the $\hat{z}$ axis, $\left\langle
F_{z}\right\rangle $ does not precess and  the Bell-Bloom
magnetometer has a dead zone for $\mathbf{B}$ parallel to the light
direction. For this case, however, one can see that the alignment
$\left\langle F_{x}^{2}-F_{y}^{2}\right\rangle $ will precess at
$2\omega _{L}$ (Fig. 1B). Hence it can be synchronously excited by
modulation in $S_{x}$ when $\omega =\omega _{L}$ and the
transmission through the cell will increase due to the
$S_{x}(F_{x}^{2}-F_{y}^{2})$ $\propto -\cos ^{2}(2\omega t)$ term.
Thus, average transmission through the cell will exhibit a CPT
transmission resonance at the same frequency $\omega =\omega _{L}$
for all orientations of the magnetic field.

The experimental apparatus and the atomic levels diagrams are
presented in Fig.~\ref{fig 1:Experimental setup and level diagram}.
A 795 nm DFB (Distributed Feedback) diode laser is used for
excitation and detection of the D1 transition in $^{87}$Rb. A linear
polarizer and an Electro-Optic Modulator (EOM) with its principle
axes at $45^{\circ }$ degrees relative to the input polarization are
used for generating the time dependent polarization modulation seen
in Fig. \ref{fig 1:Experimental setup and level diagram}. The
modulated light is sent through a $ 2\times 2\times 5$ cm$^{3}$
glass cell with isotopically enriched $^{87}$Rb metal and 10 Torr of
N$_{2}$ buffer gas. The walls of the cell are coated with OTS
coating and allow about 800 bounces before spin relaxation
\cite{SelterCoating}. The cell is heated to $78^{\circ}$C and
located inside 3 layers of $\mu $ metal shielding, achieving
magnetic field screening of $\sim 4$ orders of magnitude. A set of
coils insides the shields is used for applying the needed magnetic
fields in any orientation. Another set of coils was used for
nullifying residual magnetic field gradients. The transmitted light
intensity through the cell is measured with a photo-detector.

The wavelength dependance of the vector and tensor interactions for
$^{87}$Rb vapor in the presence of buffer gas and Doppler broadening
has been considered in \cite{MathurTangHapperPRA1970}. The imaginary
part of $H$ is maximized for both vector and tensor components close
to the $F=2$ $\rightarrow $ $F^{\prime}=1$ D1 transition. The
maximum of the scalar absorption also occurs near the same
frequency, hence the laser frequency can be locked to the
transmission minimum using laser frequency modulation.

\begin{figure}[h]
\includegraphics[width=3.3in] {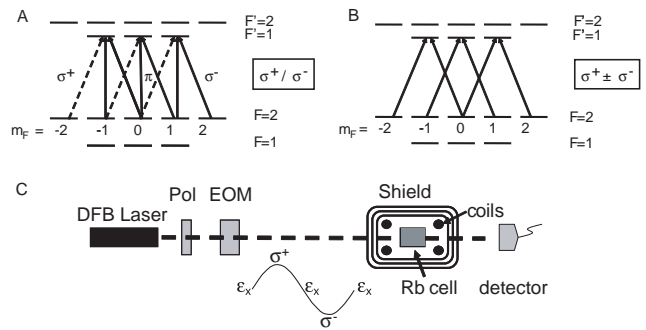}
\centering \caption{Energy levels diagrams for Bell-Bloom
magnetometer with $\sigma ^{+}$ or $\sigma ^{-}$ polarized light
(A), CPT magnetometer with linear $\sigma ^{+}\pm\sigma ^{-}$ light
(B), and the experimental apparatus (C).} \label{fig 1:Experimental
setup and level diagram}
\end{figure}

\begin{figure}[h]
\centering
\includegraphics[width=2.8in] {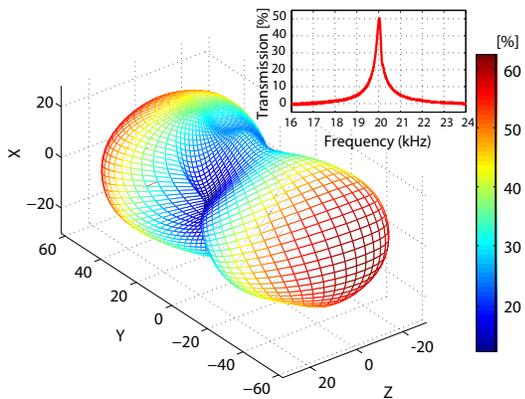}
\caption{Contrast measurements as a function of magnetic field
direction at 58 orientations, represented in percent of DC
transmission. An example of the measured contrast in one orientation
is shown in the inset.} \label{fig:Fig2_3D_image}
\end{figure}

The applied voltage on the EOM was set to oscillate sinusoidally
with a retardation amplitude of approximately $\phi _{0} \sim 2$ and
the modulation frequency was scanned around the Larmor frequency
$\omega _{L}$. In order to check the response of the magnetometer to
magnetic fields in any orientation, superposition of electrical
currents was injected to the three axes of the Helmholtz coils
inside the shields. An example of the measured transparency peak is
shown in the inset of Fig. \ref{fig:Fig2_3D_image}. The applied
magnetic field was 28.6 mG, which corresponds to a Larmor frequency
of 20 kHz.  The contrast in this case is 50$\%$ and the FWHM of the
CPT transmission resonance is 350~Hz. The measured contrast detected
from each trace, for equally spread 58 orientations, was used for
constructing the 3D plot in Fig. \ref{fig:Fig2_3D_image}. Cubic
spline interpolation was used for obtaining clearer 3D view of the
measured results. As shown in the 3D plot, there are no dead zones.
One can also see that the measured contrast along the three major
axes is different.

The natural axes of our scheme are set by the light propagation
direction, $\hat{k}_z$, the orientation of the linear polarization,
$\hat{\varepsilon}_x$ and the magnetic field. In the case of $B_z$
field parallel to the propagation vector, the linear polarization
component of the light modulated at 2$\omega_L$ generates CPT
between appropriate Zeeman sub-levels ($\Delta m=2$) (Fig. \ref{fig
1:Experimental setup and level diagram}B). Destructive interference
of the transition amplitudes in this $\Lambda$ system will induce a
transparency peak at resonance \cite{Wynands_1998}. The BB signal is
zero for this orientation. For a magnetic field oriented in a plane
perpendicular to the propagation vector $\hat{k}_z$ ($B_x$, $B_y$),
the circular polarization components ($\sigma^+$ or $\sigma^-$)
modulated at the Larmor frequency ($\omega_L$) will induce a BB
transparency peak at the Zeeman resonance ($\Delta m=1$, Fig.
\ref{fig 1:Experimental setup and level diagram}A). With the
quantization axis in the transverse plane the circularly polarized
light can be thought of as ``$\pi$'' polarization and ``$\sigma^+
\pm \sigma^-$'' polarization generating a transparency peak
resonance at $\omega_L$. This orientation term will give a symmetric
contribution to the transparency along $B_x$ and $B_y$ axes. In
addition, the CPT resonance at 2$\omega_L$ (alignment term) will
also contribute to the signal in this plane. One can see from Eq.
(\ref{eq:Hamiltonian}) that there are two dead zones in CPT
resonance for magnetic field in $\pm\hat{x}\pm \hat{y}$ directions,
as expected for a rank-2 tensor. For B field at $\pm45^{\circ}$ to
$\hat{x}$, the signal is entirely due to BB resonance. When both CPT
and BB signals are present, their interaction is complicated by the
fact that large spin orientation created by BB modulation also leads
to a significant alignment. The sign of this contribution to the
second term in Eq. (\ref{eq:Hamiltonian}) is opposite for magnetic
field in $\hat{x}$ and $\hat{y}$ directions. As a result, CPT and BB
resonances add constructively for $B_y$ and destructively for $B_x$.
By choosing an appropriate modulation depth and laser power one can
adjust the relative strength of the two signals and make the ratio
of maximum to minimum contrast to be approximately 3:1 while
maximizing the overall contrast. Experimentally we find the measured
contrast ranges between 15-60$\%$ while the FWHM is 350-700Hz,
without any dead zones, as seen in Fig.~\ref{fig:Fig2_3D_image}.

\begin{figure}[h]
\includegraphics[width=3.4in,height=!] {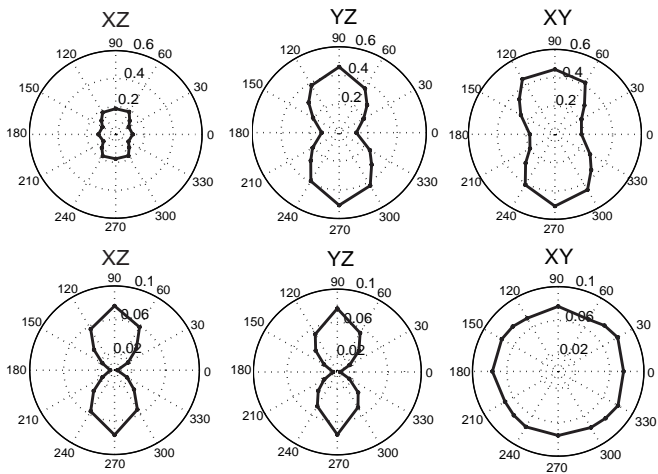}
\centering \caption{Contrast measurements along the three major
planes. Upper set corresponds to the new magnetometer. The lower set
corresponds to a specific laser light detuning (see text) in which
the CPT contribution to the signal is canceled and only ``regular''
Bell-Bloom contribution appears with a clear Dead Zone along the
$\hat{z}$ axis.  Note the factor of 6 difference in scales between
the two sets and that the chosen laser detuning for canceling the
CPT  signal is not necessarily an optimized case for the BB
contrast.} \label{fig:Fig 3 contrast 3 axes}
\end{figure}

In Fig. \ref{fig:Fig 3 contrast 3 axes} detailed contrast
measurements along the three major planes are shown. The upper set
corresponds to the signal with both CPT and BB signals, with
parameters similar to data in Fig. \ref{fig:Fig2_3D_image}. The
lower set is a reference measurement where the CPT contribution to
the signal is eliminated by choosing a specific laser detuning to
the midpoint between $F^{\prime}=1,2$ upper levels where the
constructive interference of the CPT signal is exactly cancelled due
to different phase contribution of the two upper levels. It is
evident from the figure that no signal is measured along the
$\hat{z}$ axis in the BB scheme whereas this "dead zone" is totaly
avoided in the new scheme. Moreover, along one axis ($\hat{y}$), the
signal's strength is much larger and narrower than common BB
magnetometers' signal, demonstrating a better signal to noise ratio.

The fast and balanced alternation between right and left circular
polarization at the Larmor frequency also eliminates the problem
with heading errors. Heading error induced by an asymmetric
illumination of the atoms is manifested at magnetic fields on the
order of the Earth's magnetic field ($\sim$0.5Gauss) due to the
non-linear terms in the Breit-Rabi formula. In order to check this
issue, we used the same experimental setup at a higher magnetic
field. The EOM was driven at 1.9-2.2 MHz, corresponding to about 3
Gauss for $^{87}$Rb atoms. By working at a field which is $\sim$6
times the Earth's magnetic field any distortion in the transparency
due to the quadratic nature of the Breit-Rabi splitting should be
observed.

\begin{figure}[h]
\centering
\includegraphics[width=3.3in,height=!] {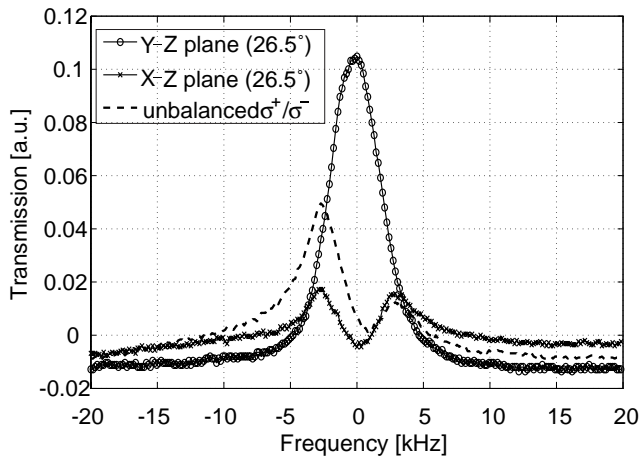}
\caption{The magnetometer's response at 2.85 Gauss (2 MHz). The two
symmetric traces correspond to a magnetic field tilted toward the
$\hat{z}$ axis by 26.5$^{\circ}$ from  the $\hat{x}$ or the
$\hat{y}$ axis. The asymmetric signal (in the X-Z plane) corresponds
to an unbalanced $\sigma^+ / \sigma^-$ contribution, obtained by
inducing an additional bias to the EOM's driving voltage.}
\label{Fig_4_Heading_Error_free}
\end{figure}

The measured signal in Fig.~\ref{Fig_4_Heading_Error_free}
corresponds to a case where a 2.85 Gauss magnetic field is tilted toward the Z axis by 26.5$%
^{\circ}$ from either the $\hat{x}$ \emph{or} $\hat{y}$ axis. This
compares signals with the largest possible contrast difference,
corresponding to the sum and the difference of the BB and CPT
contributions. Whereas at low fields a single narrow peak is
observed (e.g. Fig. \ref{fig:Fig2_3D_image}), in this large field
regime the Zeeman resonance is split into multiple peaks,
particularly pronounced in the X-Z plane. The crucial observation
here is that the signals are \emph{centered} and \emph{symmetric}.
Therefore, changes between the strength of BB and CPT contributions
associated with rotation of the magnetometer in an external field
will not result in a shift of the central frequency.

In Fig.~\ref{Fig_4_Heading_Error_free} we also present a case where
unbalanced $\sigma^+ / \sigma^-$ contribution in the modulated light
field is obtained by inducing an additional bias to the EOM's
driving voltage. In this case the asymmetric signal will induce a
heading error. This error is  avoided in the current scheme due to
the symmetric contribution of the various polarization components.
In practice, one would also have to minimize residual polarization
bias from stress-induced birefrigence in various optical elements to
obtain good suppression of heading errors.

In summary, a fundamental problem of ``dead zones'' in atomic
magnetometers caused by the vectorial or tensorial interaction with
the light has been resolved by a simple polarization modulation
scheme. Such technique can be applied to both alkali-metal and
metastable $^4$He magnetometers. We expect that the sensitivity of
such magnetometer can reach picotesla level since measured CPT
contrast ratio and linewidth are similar or better than in previous
CPT magnetometers \cite{Wynands2001,Kitching}. In the simplest
implementation of the magnetometer, frequency modulation at a few
hundred hertz of the voltage applied to the EOM can be used to lock
to the maximum of the average transmission through the cell.  It is
also possible to further improve the performance of the magnetometer
by measuring the phase of the second and fourth harmonics of the
Larmor frequency in the transmission signal, analyzing the
polarization of the transmitted light or using a separate probe
laser. The amplitude of the EOM excitation can also be continuously
adjusted to maximize the signal for any given orientation of the
magnetic field.

The heading errors specific to alkali-metal magnetometers in
geomagnetic field are also largely eliminated due to symmetric
optical pumping. Thus we expect that this technique will enable high
precision isotropic magnetometry in many Earth-bound and space
applications. Moreover, controlled excitation of orientation and
alignment by polarization modulation of light can be used for better
control of the atomic density matrix in various atomic physics
experiments.

We would like to thank Vishal Shah for assistance with the
measurements and Tom Kornack and Scott Seltzer for fabricating the
cell. This work was supported by an ONR MURI award.

\bibliographystyle{apsrev}
\bibliography{references_11}

\begin{thebibliography}{25}
\expandafter\ifx\csname natexlab\endcsname\relax\def\natexlab#1{#1}\fi
\expandafter\ifx\csname bibnamefont\endcsname\relax
  \def\bibnamefont#1{#1}\fi
\expandafter\ifx\csname bibfnamefont\endcsname\relax
  \def\bibfnamefont#1{#1}\fi
\expandafter\ifx\csname citenamefont\endcsname\relax
  \def\citenamefont#1{#1}\fi
\expandafter\ifx\csname url\endcsname\relax
  \def\url#1{\texttt{#1}}\fi
\expandafter\ifx\csname urlprefix\endcsname\relax\def\urlprefix{URL }\fi
\providecommand{\bibinfo}[2]{#2}
\providecommand{\eprint}[2][]{\url{#2}}

\bibitem[{\citenamefont{Budker and Romalis}(2007)}]{Budker_Romalis_Nature_Phys}
\bibinfo{author}{\bibfnamefont{D.}~\bibnamefont{Budker}} \bibnamefont{and}
  \bibinfo{author}{\bibfnamefont{M.}~\bibnamefont{Romalis}},
  \bibinfo{journal}{Nature Phys.} \textbf{\bibinfo{volume}{3}},
  \bibinfo{pages}{227} (\bibinfo{year}{2007}).

\bibitem[{\citenamefont{Dougherty et~al.}(2004)}]{Cassini}
\bibinfo{author}{\bibfnamefont{M.}~\bibnamefont{Dougherty}}
  \bibnamefont{et~al.}, \bibinfo{journal}{Space Science Rev.}
  \textbf{\bibinfo{volume}{114}}, \bibinfo{pages}{331} (\bibinfo{year}{2004}).

\bibitem[{\citenamefont{Matousek}(2007)}]{Juno}
\bibinfo{author}{\bibfnamefont{S.}~\bibnamefont{Matousek}},
  \bibinfo{journal}{Acta Astronautica} \textbf{\bibinfo{volume}{61}},
  \bibinfo{pages}{932} (\bibinfo{year}{2007}).

\bibitem[{\citenamefont{Friis-Christensen
  et~al.}(2006)\citenamefont{Friis-Christensen, L\"{u}hr, and Hulot}}]{Swarm}
\bibinfo{author}{\bibfnamefont{E.}~\bibnamefont{Friis-Christensen}},
  \bibinfo{author}{\bibfnamefont{H.}~\bibnamefont{L\"{u}hr}}, \bibnamefont{and}
  \bibinfo{author}{\bibfnamefont{G.}~\bibnamefont{Hulot}},
  \bibinfo{journal}{Earth Planets Space} \textbf{\bibinfo{volume}{58}},
  \bibinfo{pages}{351} (\bibinfo{year}{2006}).

\bibitem[{\citenamefont{Nabighian et~al.}(2005)}]{Nabighian_2005}
\bibinfo{author}{\bibfnamefont{M.}~\bibnamefont{Nabighian}}
  \bibnamefont{et~al.}, \bibinfo{journal}{Geophysics}
  \textbf{\bibinfo{volume}{70}}, \bibinfo{pages}{33ND} (\bibinfo{year}{2005}).

\bibitem[{\citenamefont{David et~al.}(2004)}]{David_2004}
\bibinfo{author}{\bibfnamefont{A.}~\bibnamefont{David}} \bibnamefont{et~al.},
  \bibinfo{journal}{Antiquity} \textbf{\bibinfo{volume}{78}},
  \bibinfo{pages}{341} (\bibinfo{year}{2004}).

\bibitem[{\citenamefont{Nelson et~al.}(2005)\citenamefont{Nelson, McDonald, and
  Wright}}]{Nelson}
\bibinfo{author}{\bibfnamefont{H.}~\bibnamefont{Nelson}},
  \bibinfo{author}{\bibfnamefont{J.}~\bibnamefont{McDonald}}, \bibnamefont{and}
  \bibinfo{author}{\bibfnamefont{D.}~\bibnamefont{Wright}},
  \bibinfo{type}{Tech. Rep.} \bibinfo{number}{NRL/MR/6110--05-8874},
  \bibinfo{institution}{NRL} (\bibinfo{year}{2005}).

\bibitem[{\citenamefont{Bloom}(1962)}]{Bloom_APL_1962}
\bibinfo{author}{\bibfnamefont{A.~L.} \bibnamefont{Bloom}},
  \bibinfo{journal}{Appl. Opt.} \textbf{\bibinfo{volume}{1}},
  \bibinfo{pages}{61} (\bibinfo{year}{1962}).

\bibitem[{\citenamefont{Cheron et~al.}(1997)}]{Cheron_J_Phys_III_1997}
\bibinfo{author}{\bibfnamefont{B.}~\bibnamefont{Cheron}} \bibnamefont{et~al.},
  \bibinfo{journal}{J. Phys. III France} \textbf{\bibinfo{volume}{7}},
  \bibinfo{pages}{1735} (\bibinfo{year}{1997}).

\bibitem[{\citenamefont{Ch\'eron et~al.}(2001)\citenamefont{Ch\'eron, Gilles,
  and Hamel}}]{Cheron2000}
\bibinfo{author}{\bibfnamefont{B.}~\bibnamefont{Ch\'eron}},
  \bibinfo{author}{\bibfnamefont{H.}~\bibnamefont{Gilles}}, \bibnamefont{and}
  \bibinfo{author}{\bibfnamefont{J.}~\bibnamefont{Hamel}},
  \bibinfo{journal}{Eur. Phys. J. AP} \textbf{\bibinfo{volume}{13}},
  \bibinfo{pages}{143} (\bibinfo{year}{2001}).

\bibitem[{\citenamefont{Guttin et~al.}(1994)\citenamefont{Guttin, Leger, and
  Stoeckel}}]{Guttin_J_De_Phys_1997}
\bibinfo{author}{\bibfnamefont{C.}~\bibnamefont{Guttin}},
  \bibinfo{author}{\bibfnamefont{J.}~\bibnamefont{Leger}}, \bibnamefont{and}
  \bibinfo{author}{\bibfnamefont{F.}~\bibnamefont{Stoeckel}},
  \bibinfo{journal}{J. Phys. IV Collocq France} \textbf{\bibinfo{volume}{55}},
  \bibinfo{pages}{C4 665} (\bibinfo{year}{1994}).

\bibitem[{\citenamefont{Aleksandrov et~al.}(2006)\citenamefont{Aleksandrov,
  Vershovskii, and Pazgalev}}]{Aleksandrov2006}
\bibinfo{author}{\bibfnamefont{E.~B.} \bibnamefont{Aleksandrov}},
  \bibinfo{author}{\bibfnamefont{A.~K.} \bibnamefont{Vershovskii}},
  \bibnamefont{and} \bibinfo{author}{\bibfnamefont{A.~S.}
  \bibnamefont{Pazgalev}}, \bibinfo{journal}{Technical Physics}
  \textbf{\bibinfo{volume}{51}}, \bibinfo{pages}{919–} (\bibinfo{year}{2006}).

\bibitem[{\citenamefont{Bell and Bloom}(1957)}]{Bell_Bloom_1957}
\bibinfo{author}{\bibfnamefont{W.~E.} \bibnamefont{Bell}} \bibnamefont{and}
  \bibinfo{author}{\bibfnamefont{A.~L.} \bibnamefont{Bloom}},
  \bibinfo{journal}{Phys. Rev.} \textbf{\bibinfo{volume}{107}},
  \bibinfo{pages}{1559} (\bibinfo{year}{1957}).

\bibitem[{\citenamefont{Gilles et~al.}(1991)\citenamefont{Gilles, Cheron, and
  Hamel}}]{Gilles_Opt_Comm_1991}
\bibinfo{author}{\bibfnamefont{H.}~\bibnamefont{Gilles}},
  \bibinfo{author}{\bibfnamefont{J.}~\bibnamefont{Cheron}}, \bibnamefont{and}
  \bibinfo{author}{\bibfnamefont{J.}~\bibnamefont{Hamel}},
  \bibinfo{journal}{Opt. Comm} \textbf{\bibinfo{volume}{61}},
  \bibinfo{pages}{369} (\bibinfo{year}{1991}).

\bibitem[{\citenamefont{Klepel and Suter}(1992)}]{Suter}
\bibinfo{author}{\bibfnamefont{H.}~\bibnamefont{Klepel}} \bibnamefont{and}
  \bibinfo{author}{\bibfnamefont{D.}~\bibnamefont{Suter}},
  \bibinfo{journal}{Opt. Comm.} \textbf{\bibinfo{volume}{90}},
  \bibinfo{pages}{46 } (\bibinfo{year}{1992}).

\bibitem[{\citenamefont{Budker et~al.}(2002)}]{Buder_RMP_2002_NMOR}
\bibinfo{author}{\bibfnamefont{D.}~\bibnamefont{Budker}} \bibnamefont{et~al.},
  \bibinfo{journal}{Rev. Mod. Phys.} \textbf{\bibinfo{volume}{74}},
  \bibinfo{pages}{1153} (\bibinfo{year}{2002}).

\bibitem[{\citenamefont{Nagel et~al.}(1998)}]{Wynands_1998}
\bibinfo{author}{\bibfnamefont{A.}~\bibnamefont{Nagel}} \bibnamefont{et~al.},
  \bibinfo{journal}{Europhys. Lett.} \textbf{\bibinfo{volume}{44}},
  \bibinfo{pages}{31} (\bibinfo{year}{1998}).

\bibitem[{\citenamefont{Stahler et~al.}(2001)}]{Wynands2001}
\bibinfo{author}{\bibfnamefont{M.}~\bibnamefont{Stahler}} \bibnamefont{et~al.},
  \bibinfo{journal}{Europhys. Lett.} \textbf{\bibinfo{volume}{54}},
  \bibinfo{pages}{323} (\bibinfo{year}{2001}).

\bibitem[{\citenamefont{Schwindt et~al.}(2004)}]{Kitching}
\bibinfo{author}{\bibfnamefont{P.~D.} \bibnamefont{Schwindt}}
  \bibnamefont{et~al.}, \bibinfo{journal}{Appl. Phys. Lett.}
  \textbf{\bibinfo{volume}{85}}, \bibinfo{pages}{6409} (\bibinfo{year}{2004}).

\bibitem[{\citenamefont{Seltzer et~al.}(2007)\citenamefont{Seltzer, Meares, and
  Romalis}}]{Seltzer_2007}
\bibinfo{author}{\bibfnamefont{S.~J.} \bibnamefont{Seltzer}},
  \bibinfo{author}{\bibfnamefont{P.~J.} \bibnamefont{Meares}},
  \bibnamefont{and} \bibinfo{author}{\bibfnamefont{M.~V.}
  \bibnamefont{Romalis}}, \bibinfo{journal}{Phys. Rev. A}
  \textbf{\bibinfo{volume}{75}}, \bibinfo{pages}{051407(R)}
  (\bibinfo{year}{2007}).

\bibitem[{\citenamefont{Happer}(1972)}]{Happer_RMP_1972}
\bibinfo{author}{\bibfnamefont{W.}~\bibnamefont{Happer}},
  \bibinfo{journal}{Rev. Mod. Phys.} \textbf{\bibinfo{volume}{44}},
  \bibinfo{pages}{169} (\bibinfo{year}{1972}).

\bibitem[{\citenamefont{Geremia et~al.}(2006)\citenamefont{Geremia, Stockton,
  and Mabuchi}}]{Mabuchi_2006}
\bibinfo{author}{\bibfnamefont{J.~M.} \bibnamefont{Geremia}},
  \bibinfo{author}{\bibfnamefont{J.~K.} \bibnamefont{Stockton}},
  \bibnamefont{and} \bibinfo{author}{\bibfnamefont{H.}~\bibnamefont{Mabuchi}},
  \bibinfo{journal}{Phys. Rev. A} \textbf{\bibinfo{volume}{73}},
  \bibinfo{pages}{042112} (\bibinfo{year}{2006}).

\bibitem[{\citenamefont{Happer and Mathur}(1967)}]{HapperMathurPR1967}
\bibinfo{author}{\bibfnamefont{W.}~\bibnamefont{Happer}} \bibnamefont{and}
  \bibinfo{author}{\bibfnamefont{B.~S.} \bibnamefont{Mathur}},
  \bibinfo{journal}{Phys. Rev.} \textbf{\bibinfo{volume}{163}},
  \bibinfo{pages}{12} (\bibinfo{year}{1967}).

\bibitem[{\citenamefont{Mathur et~al.}(1970)\citenamefont{Mathur, Tang, and
  Happer}}]{MathurTangHapperPRA1970}
\bibinfo{author}{\bibfnamefont{B.~S.} \bibnamefont{Mathur}},
  \bibinfo{author}{\bibfnamefont{H.~Y.} \bibnamefont{Tang}}, \bibnamefont{and}
  \bibinfo{author}{\bibfnamefont{W.}~\bibnamefont{Happer}},
  \bibinfo{journal}{Phys. Rev. A} \textbf{\bibinfo{volume}{2}},
  \bibinfo{pages}{648} (\bibinfo{year}{1970}).

\bibitem[{\citenamefont{Seltzer and Romalis}(2009)}]{SelterCoating}
\bibinfo{author}{\bibfnamefont{S.~J.} \bibnamefont{Seltzer}} \bibnamefont{and}
  \bibinfo{author}{\bibfnamefont{M.~V.} \bibnamefont{Romalis}},
  \bibinfo{journal}{J. Appl. Phys.} \textbf{\bibinfo{volume}{106}},
  \bibinfo{pages}{114905} (\bibinfo{year}{2009}).

\end{thebibliography}

\end{document}